\documentclass[a4paper]{article}
\usepackage{xcolor}
\usepackage{graphicx}
\usepackage{hyperref}
\usepackage{subcaption}
\usepackage{amsmath,amssymb,latexsym}
\usepackage{amsfonts}
\title{
\begin{flushright}
{\small INR-TH-2018-010}
\end{flushright}
Vibrational modes of Q-balls}

\author{A.\;Kovtun$^{a,b}$\thanks{{\bf e-mail}:
andrei.kovtun@phystech.edu}, E.\;Nugaev$^{a,b}$\thanks{{\bf e-mail}:
emin@ms2.inr.ac.ru}, A.\;Shkerin$^{c,a}$\thanks{{\bf e-mail}:
andrey.shkerin@epfl.ch}
\\
$^a${\small{\em
Institute for Nuclear Research of the Russian Academy
of Sciences,}}\\
{\small{\em 60th October Anniversary prospect 7a, 117312, Moscow,
Russia
}}\\
$^b${\small{\em Moscow Institute of Physics and Technology,
}}\\
{\small{\em Institutskii per. 9, Dolgoprudny, Moscow Region 141700,
Russia
}}\\
$^c${\small{\em Institute of Physics, Ecole Polytechnique F\'ed\'erale de Lausanne (EPFL),
}}\\
{\small{\em  CH-1015, Lausanne, Switzerland}}\\
}

\date{}

\begin{document}

\maketitle

\begin{abstract}
We study linear perturbations of classically stable Q-balls in theories admitting analytic solutions. Although the corresponding boundary value problem is non-Hermitian, the analysis of perturbations can also be performed analytically in certain regimes. We show that in theories with the flat potential, large Q-balls possess soft excitations. We also find a specific vibrational mode for Q-balls with a near-critical charge, where the perturbation theory for excitations can be developed. Comparing with the results on stability of Q-balls provides additional checks of our analysis.
\end{abstract}

\section{Introduction}

The study of linear perturbations above compact objects is a general issue which can provide interesting phenomenology. Recent examples include the investigation of quasinormal modes of black holes, which play an essential role in the analysis of gravitational wave signals \cite{Konoplya:2016pmh,Cardoso:2016oxy}. For stars composed of baryonic matter, the perturbations depend strongly on the internal physics and cannot be surveyed by the means of field theory. Moreover, the presence of dissipation usually leaves possible only numerical analysis in a realistic setting. 

To avoid problems related to dissipation, one can study closed systems that can be formulated in the Lagrangian form. Nonrelativistic physics provides a very interesting possibility to investigate a nonuniform Bose-Einstein condensate, see \cite{Dalfovo:1999zz} for a review. It is remarkable that excitations in the Bose-Einstein condensate can be studied both theoretically and experimentally in the dilute gas approximation. Relativistic field theories of
a complex scalar field also provide compact objects that can be studied in the semiclassical
approximation.
Examples of localized stationary configurations with dynamical gravity include different types of boson stars (BS) (see \cite{Schunck:2003kk,Liebling:2012fv} for reviews) or related objects such as axion miniclusters \cite{Kolb:1993zz}.  Recently, the analysis of bound states of BS was performed in \cite{Eby:2017teq}. Although their phenomenology is interesting, gravity precludes the analytical investigation of excitations of BS. Objects better suited for this purpose are counterparts of (solitonic) BS in the limit when gravity is absent. These are Q-balls \cite{Coleman:1985ki} arising in theories of a complex scalar field with potentials of a special type.  

In this paper we first discuss applicability conditions for the semiclassical treatment of solitons (section \ref{sec:appr}). We then study linear perturbations of Q-balls in a piecewise parabolic potential with the flat direction. In this model, the perturbations can be found in an analytic form at all frequencies of the background configuration. This is done in section \ref{sec:PPP}. In section \ref{sec:Poly}, we consider perturbations in the thin-wall limit of Q-balls in a polynomial potential. One can think of this case as a complement to the model with the flat potential where the thin-wall approximation is not applicable. In appendix \ref{sec:log} we use a model with an analytic solution to check the perturbation theory near the critical charge, developed in the main text.

\section{Semiclassical approximation}
\label{sec:appr}

Consider a $U(1)$-invariant theory of the complex scalar field with the Lagrangian
\begin{equation}
\mathcal{L}=\left|\partial_{\mu}\phi\right|^{2}-V(|\phi|) \; .
\end{equation}
The potential $V(|\phi|)$ is assumed to be a smooth function of $|\phi |^2$ with an absolute minimum at
$|\phi |=0$ at which $V(0)=0$. For the validity of the semiclassical description of solitons \cite{Lee:1991ax}, the potential must be presentable in the form
\begin{equation}\label{appr:U}
V(|\phi|)=\frac{1}{g^2}U(g |\phi |) \; ,
\end{equation}
where $g$ is a small dimensionless coupling constant. Then, after an appropriate field redefinition the action of the theory becomes supplemented with the overall large factor $g^{-2}$, thus justifying the semiclassical treatment of solitons. The form (\ref{appr:U}) of the potential is convenient in $(1+1)$ dimensions where the field $\phi$ is itself dimensionless. In four-dimensional space-time let us suppose that the relevant scale of a theory is set by a mass $m$ of the boson in the vacuum $\phi=0$. Then, eq. (\ref{appr:U}) is rewritten as
\begin{equation}\label{appr:U4}
V(|\phi|)=\frac{m^4}{g^2}U(g|\phi|/m) \; .
\end{equation}
 
In this paper we consider two featured potentials giving essentially different profiles for classically stable Q-balls. The first one is the piecewise parabolic potential \cite{Rosen:1968}, 
\begin{equation}\label{appr:PPP}
V(|\phi|)=m^2 |\phi|^2 \theta\left( 1-\frac{|\phi|^2}{v^2} \right) + m^2 v^2 \theta\left( \frac{|\phi|^2}{v^2} - 1\right) \; ,
\end{equation}
which admits analytic solutions both for the classical configurations and linear perturbations above them \cite{Gulamov:2013ema}. It can be written in the form (\ref{appr:U4}) with $g=m/v$ and
\begin{equation}\label{appr:PPP_U}
U(x)=x^2\theta(1-x^2)+\theta(x^2-1) \; .
\end{equation}
The potential (\ref{appr:PPP}) serves to approximate realistic potentials with the flat direction, which can be of use, e.g., in describing dark matter by Q-balls \cite{Kusenko:1997si}. The Heaviside functions in eq. (\ref{appr:PPP_U}) can be regularized in a number of ways by replacing
\begin{equation}
U(x)\to U_\alpha(x) \; ,
\end{equation}
where we assume that the regularization parameter $\alpha$ is not very large and
\begin{equation}
g \ll \alpha \; .
\end{equation}
Other theories admitting nontopological solitons contain potentials of the polynomial form. As was discussed in \cite{Coleman:1985ki}, in order to allow Q-balls in a theory of one complex scalar field with the global $U(1)$-symmetry, it is necessary to include nonrenormalizable self-interactions into the polynomial scalar potential. Yet, it is enough to add just the sixth-order term $\propto | \phi |^6$. In \cite{Mai:2012yc}, the thorough analysis of Q-balls in the polynomial potential of the sixth degree was performed. For practical purposes, one can absorb the constants $m$ and $g$ in a suitable field redefinition, by writing $m=g=1$. Of course, this does not mean strong coupling and the overall factor $g^{-2}$ in the action must be kept in mind.

The ansatz for spherically symmetric Q-balls takes the form
\begin{equation}\label{appr:Qball}
\phi_0(\vec{x},t)=f(r)e^{i\omega t} \; .
\end{equation}
The conserved $U(1)$ charge is then given by
\begin{equation}
Q=-i\int d^3x (\phi_0^\ast\dot{\phi}_0-\phi_0\dot{\phi}_0^\ast)=8\pi\omega\int dr r^2 f^2(r) \; ,
\end{equation}
and we will limit the consideration to the case $Q>0$, which implies the positive frequencies, $\omega>0$. The charge conservation does not guarantee stability of nontopological solitons. The latter, therefore, split on two branches --- the classically stable Q-balls and the classically unstable ones, dubbed as Q-clouds \cite{Alford:1987vs}.\footnote{Classically stable Q-balls can still be metastable in the small kinematical region; see \cite{Levkov:2017paj} for details. }

We will study small perturbations of the configurations (\ref{appr:Qball}),
\begin{equation}
\phi=\phi_0+\chi \; , ~~~~ \chi(\vec{x},t)=\psi(\vec{x},t) {\it e}^{i\omega t} \; ,
\end{equation}
where we factored out the background phase factor in the expression for $\chi$ in order to get rid of the explicit time dependence in the linearized equations for perturbations. The latter reads as follows:
\begin{align}\label{appr:psi}
\begin{split}
& \left(\partial_{0}+i \omega\right)^{2}\psi -\Delta\psi= -h(r)\psi^{\ast}-g(r)\psi \; , \\ 
& \left(\partial_{0}-i \omega\right)^{2} \psi^{\ast} -\Delta\psi^{\ast}= -h(r)\psi-g(r)\psi^{\ast} \; ,
\end{split}
\end{align}
where
\begin{align}\label{appr:g,h}
\begin{split}
h(r) & =\left.\left(z\frac{d^2V}{dz^2}\right)\right|_{z=f^2(r)} \; ,\\
g(r) & =\left.\left(z\frac{d^2V}{dz^2}+\frac{dV}{dz}\right)\right|_{z=f^2(r)} \; , 
\end{split}
\end{align}
and we denoted $z=|\phi|^2$. Note that, in general, it is impossible to disentangle the equations for $\psi$ and its complex conjugate. Also, eqs. (\ref{appr:psi}) cannot be viewed as an eigenvalue problem for some Hermitian operator. We will see how to deal with this fact in the examples below.

\section{Perturbations in Piecewise parabolic potential}
\label{sec:PPP}

In this section we study perturbations of Q-balls in three spatial dimensions and in a theory with the piecewise parabolic potential eq. (\ref{appr:PPP}). For this potential, the functions $h$, $g$ defined in eqs. (\ref{appr:g,h}) become
\begin{align}\label{PPP:h,g}
\begin{split}
h(r) & =-\frac{m^2}{2}\delta\left( \frac{f(r)}{v}-1 \right) \; \\
g(r) & =m^2 \theta\left( 1-\frac{f^2(r)}{v^2} \right)+h(r) \; .
\end{split}
\end{align}
Our main concern will be with the discrete spectrum of modes around a classically stable Q-ball solution whose frequency is below the critical one, $\omega<\omega_c$. We will refer to them as the vibrational modes. For completeness, we also consider a decay mode of the unstable branch of Q-balls, occurring at $\omega>\omega_c$. As we will see, an analytic continuation of the decay mode below the cusp point $\omega_c$ gives a specific vibrational mode existing for the solutions with the frequencies close enough to $\omega_c$.

From eqs. (\ref{PPP:h,g}) one observes the mass $m$ of the boson to be the relevant dimensional parameter in the theory. For this reason, in this section we measure physical quantities in the units of this mass, by setting $m=1$. Then, $\omega_c\approx 0.960$.

\subsection{Vibrational modes}
\label{ssec:vibro_modes}

An appropriate ansatz \cite{Smolyakov:2017axd} governing the dynamics of small oscillations on top of the classically stable Q-balls reads as follows:\footnote{Note that the ansatz (\ref{ppp:vibro_ansatz}) does not allow us to catch all possible excitations of a Q-ball. For example, it excludes the $U(1)$ modes from consideration unless $\gamma=0$. }
\begin{equation}\label{ppp:vibro_ansatz}
\psi(\vec{x},t)=(\psi^{(l)}_1(r)e^{i\gamma t}+\psi^{(l)}_2(r)e^{-i\gamma t})Y_{l,m}(\theta,\varphi) \; ,
\end{equation}
where the parameter $\gamma$ is taken to be real and positive, $\psi^{(l)}_1$, $\psi^{(l)}_2$ are real functions of the radial coordinate and $Y_{l,m}$ are spherical harmonics. Substituting this into eq. (\ref{appr:psi}) and collecting the terms with equal phase factors, we obtain
\begin{align}\label{ppp:vibro_pert_eqns}
\begin{split}
\left(\Delta_{r}-\frac{l(l+1)}{r^2} +(\omega+\gamma)^2-g(r)\right)\psi_{1}^{(l)}(r)-h(r)\psi_{2}^{(l)\ast}(r)=0 \; , \\
\left(\Delta_{r}-\frac{l(l+1)}{r^2} +(\omega-\gamma)^2-g(r)\right)\psi_{2}^{(l)}(r)-h(r)\psi_{1}^{(l)\ast}(r)=0 \; .
\end{split}
\end{align} 
The equations for perturbations must be supplemented with the boundary conditions
\begin{equation}\label{ppp:bc}
\psi_{1,2}^{(l)}(\infty)=0 \; , ~~~~  \left.\frac{d \psi_{1,2}^{(l)}}{d r}\right|_{r=0}=0 \; .
\end{equation}
In order to make the solution of eqs. (\ref{ppp:vibro_pert_eqns}) satisfy the boundary condition at infinity (i.e., to limit the analysis to the bound states), we have to impose
\begin{equation}\label{ppp:o+g<1}
\omega+\gamma<1 \; .
\end{equation}

Eqs. (\ref{ppp:vibro_pert_eqns}) are solved exactly in the regions of magnitudes of the background Q-ball $f(r)>v$ ($r<R$) and $f(r)<v$ ($r>R$). Namely,
\begin{equation}\label{ppp:psi_1}
\psi_1^{(l)}(r)=\left\lbrace
\begin{array}{ll}
\displaystyle A\,\frac{J_{l+1/2}(\omega_+r)}{\sqrt{r}} \; , & r<R \\
\\
\displaystyle B\,\frac{K_{l+1/2}(\lambda_+r)}{\sqrt{r}} \; , & r> R 
\end{array}\right. 
\end{equation}
and
\begin{equation}\label{ppp:psi_2}
\psi_2^{(l)}(r)=\left\lbrace
\begin{array}{ll}
\displaystyle C\,\frac{J_{l+1/2}(\omega_-r)}{\sqrt{r}} \; , & r<R \\
\\
\displaystyle D\,\frac{K_{l+1/2}(\lambda_-r)}{\sqrt{r}} \; , & r> R 
\end{array}\right. 
\end{equation}
where $J_{l+1/2}$ and $K_{l+1/2}$ are Bessel and modified Bessel functions of the first and the second kind correspondingly and of the order $l+1/2$, and we denoted
\begin{align}
\begin{split}
 \omega_{\pm} & =\omega \pm \gamma  \; ,  \\
 \lambda_{\pm} & =\sqrt{1-(\omega \pm \gamma)^2}  \; .
\end{split}
\end{align}
The solutions must be matched smoothly at the point $r=R$, which is given by
\begin{equation}\label{ppp:R}
R=\frac{1}{\omega}\left(\pi-\arctan\frac{\omega}{\sqrt{1-\omega^2}}\right) \; .
\end{equation}
This is the matching radius for the background configuration; see \cite{Gulamov:2013ema} for details. One can consider that inside this radius the scalar potential is flat and outside it is quadratic. It is this feature that allows us to resolve analytically the Q-ball profile in the potential (\ref{appr:PPP}).

The values of the constants $A$ to $D$ are inferred from the matching conditions. The latter can be resolved provided that
\begin{equation}\label{ppp:det_vib_orbital}
\left( \frac{K'_{l+1/2}(\lambda_+ R)}{K_{l+1/2}(\lambda_+ R)}-\frac{J'_{l+1/2}(\omega_+ R)}{J_{l+1/2}(\omega_+ R)}+\frac{\Lambda}{2} \right)\left( \frac{K'_{l+1/2}(\lambda_- R)}{K_{l+1/2}(\lambda_- R)}-\frac{J'_{l+1/2}(\omega_- R)}{J_{l+1/2}(\omega_- R)}+\frac{\Lambda}{2} \right)-\frac{\Lambda^2}{4}=0 \; ,
\end{equation}
where
\begin{equation}\label{ppp:L}
\Lambda=\frac{R}{R\sqrt{1-\omega^2}+1} \; .
\end{equation}
Eqs. (\ref{ppp:o+g<1}) and (\ref{ppp:det_vib_orbital}) determine the spectrum of vibrational modes near the Q-balls. At a given frequency $\omega$, there is a finite amount of modes distinguished by $l$ and an integer $n$ enumerating the solutions of eq. (\ref{ppp:det_vib_orbital}). Additionally, there is a $(2l+1)$ degeneracy in each mode with the given values of $l$ and $n$.

\begin{figure}[t]
\centering
\begin{minipage}[c]{0.49\textwidth}
\includegraphics[scale=0.3]{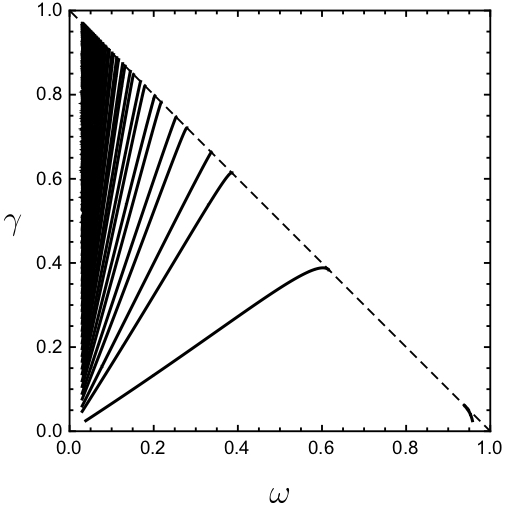}
\end{minipage}
\hfill
\begin{minipage}[c]{0.49\textwidth}
\includegraphics[scale=0.3]{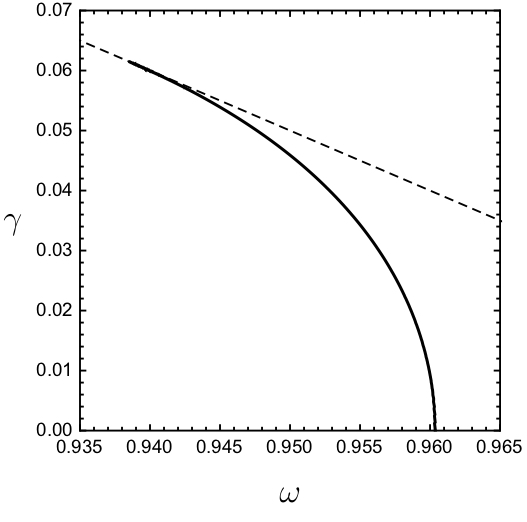}
\end{minipage}
\caption{ The discrete spectrum of linear perturbations of classically stable Q-balls in the potential (\ref{appr:PPP}), at $l=0$. The dashed line shows the bound specified by eq. (\ref{ppp:o+g<1}), which separates the discrete and continuous parts of the spectrum. All quantities are normalized to the parameter $m$. The left panel shows the overall picture, while the right panel takes a closer look at the vibrational mode near the cusp point. }
\label{fig:ppp_vibro_modes}
\end{figure}

Let us take a closer look at the spectrum of spherically symmetric modes, $l=0$. It is plotted in fig. \ref{fig:ppp_vibro_modes}. We observe that the number of bound states increases as $\omega\rightarrow 0$. In this limit, $Q\rightarrow \infty$ and the large Q-balls possess soft modes with $\gamma\rightarrow 0$; i.e., both $\gamma$ and $\omega$ are much less than the mass of the free boson. An argument in favor of this property is the flatness of the scalar potential inside the Q-balls at the scale $\sim\omega^{-1}$. Interestingly, the Q-balls with the large enough frequencies do not support any of the modes from the discrete spectrum. However, close to the stability bound one oscillatory solution reappears. As we will see below, this solution continues to the instability region where it represents the decay mode.

\begin{figure}[t]
\centering
\includegraphics[scale=0.65]{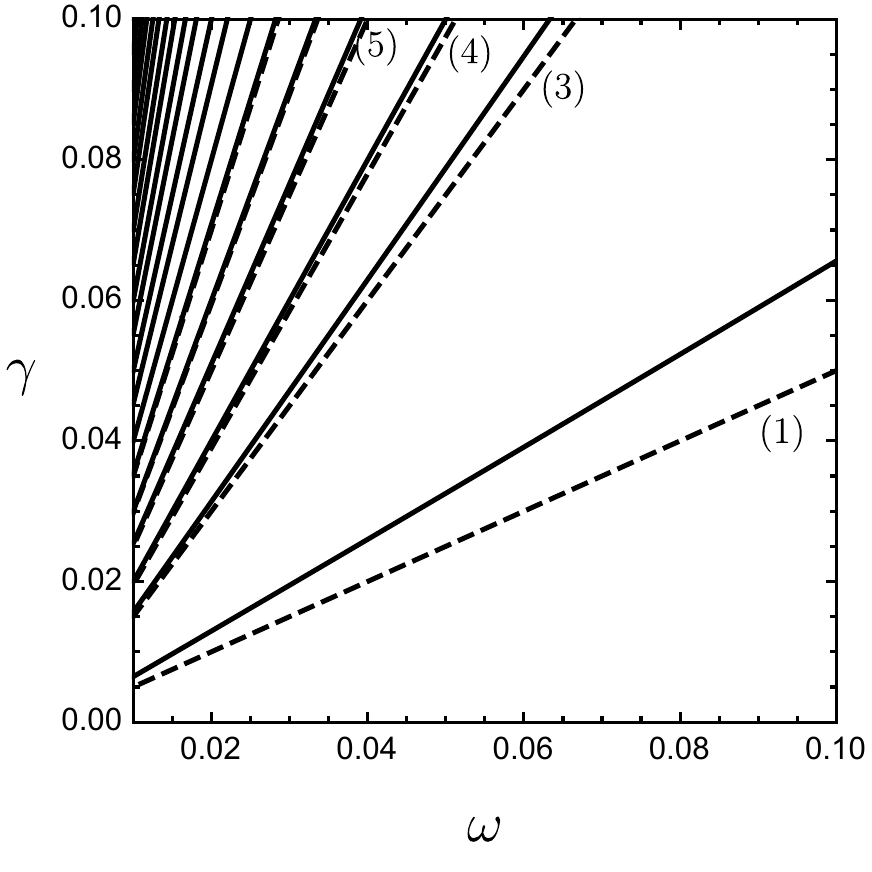}
\caption{ The discrete spectrum of spherically symmetric perturbations of Q-balls with low frequencies $\omega$. One observes the oscillation rates $\gamma$ to become proportional to $\omega$. The solid lines show the exact solution, while the dotted lines represent the approximate formula $\gamma=k\omega$ with $k$ given in eq. (\ref{ppp:k_n}). The number $(n)$ marks the mode at the $n$-th energy level, according to eq. (\ref{ppp:k_n}). }
\label{fig:ppp_linear_spectrum}
\end{figure}

From fig. \ref{fig:ppp_vibro_modes} one observes that the dependence of the relative frequency $\gamma$ of a given mode on $\omega$ linearizes in the limit $\omega\rightarrow 0$. To see this explicitly, we take $\gamma=k\omega$ and substitute it to eq. (\ref{ppp:det_vib_orbital}) with $l=0$. To the first order in $\omega$, this gives
\begin{equation}\label{ppp:k}
(k\pi\cos k\pi-\sin k\pi)\sin k\pi=0 \; .
\end{equation} 
The solution of this equation is well approximated by 
\begin{equation}\label{ppp:k_n}
k_n\approx \frac{n }{2} \; , ~~~ n=1,3,4,5,...
\end{equation}
where we have excluded the root with $n=2$ since the latter is singular; see fig. \ref{fig:ppp_linear_spectrum} for a comparison of eq. (\ref{ppp:k_n}) with the exact solution. Finally, for the vibrational mode existing in the opposite part of the spectrum, $\omega\rightarrow\omega_c$, fig. \ref{fig:ppp_vibro_modes} demonstrates that
\begin{equation}\label{ppp:near_cusp}
\gamma\sim \sqrt{\omega_c-\omega} \; .
\end{equation}

\begin{figure}[t]
\centering
\begin{minipage}[c]{0.49\textwidth}
\includegraphics[scale=0.22]{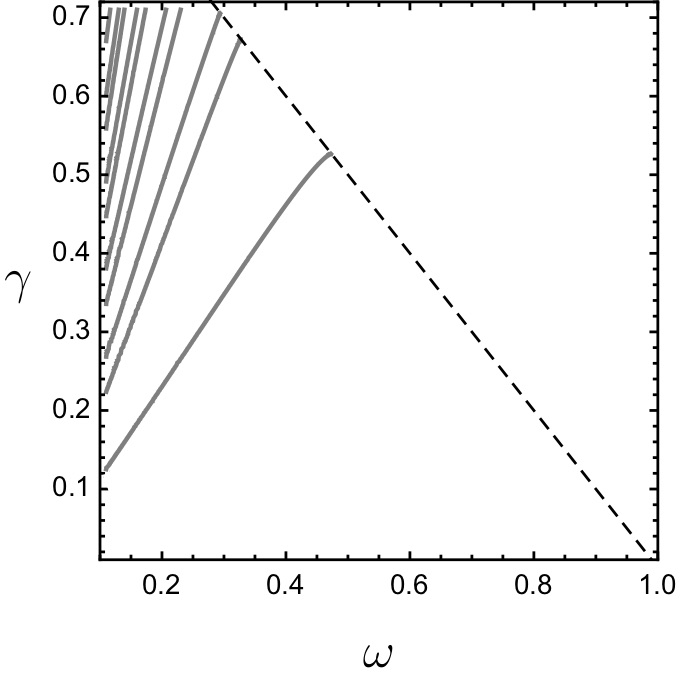}
\end{minipage}
\hfill
\begin{minipage}[c]{0.49\textwidth}
\includegraphics[scale=0.22]{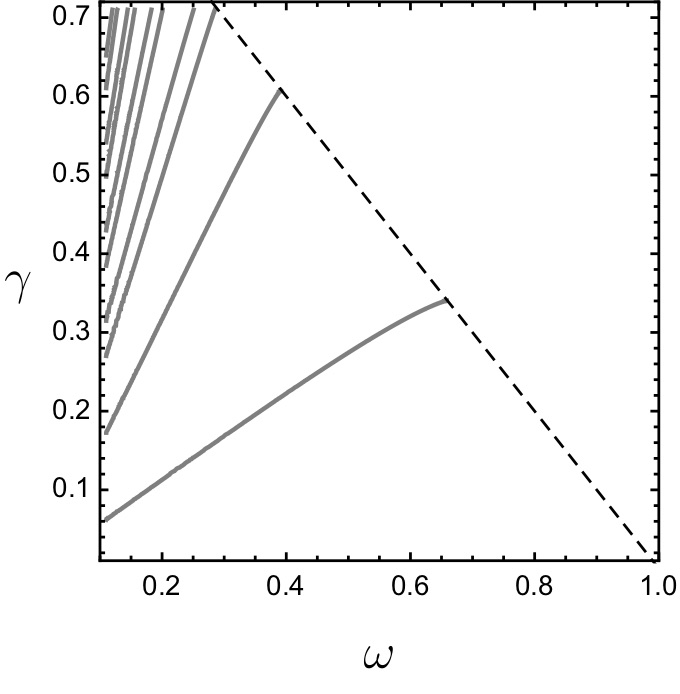}
\end{minipage}
\caption{ The discrete spectrum of linear perturbations of classically stable Q-balls in the potential (\ref{appr:PPP}), at $l=1$ (the left panel) and $l=2$ (the right panel); cf. fig. \ref{fig:ppp_vibro_modes}.  }
\label{fig:ppp_vibro_modes_l}
\end{figure}

Note that whenever $\gamma$ is small, one can make use of the perturbation theory with respect to $\gamma$. In this limit, the linear perturbations of the Q-balls take a simple form
\begin{equation}\label{ppp:psi_near_cusp_vibro}
\psi_1\sim f+\gamma\dfrac{\partial f}{\partial \omega}+\mathcal{O}(\gamma^2) \; , ~~~ \psi_2\sim -f+\gamma\dfrac{\partial f}{\partial\omega}+\mathcal{O}(\gamma^2) \; .
\end{equation}
From fig. \ref{fig:ppp_linear_spectrum} we see that, apart from the region of low frequencies, the perturbation theory suits well to describe the vibrational mode appearing when $\omega$ approaches the critical value. Note also that the expansion (\ref{ppp:psi_near_cusp_vibro}) is not peculiar to the potential (\ref{appr:PPP}). In fact, the soft modes described by the ansatz (\ref{ppp:vibro_ansatz}), as well as the mode in the vicinity of the cusp point, are of this form regardless of the shape of the scalar potential. In appendix \ref{sec:log} we explicitly illustrate this using another exactly solvable potential.

The structure of the spectrum with a nonzero orbital momentum is similar to that with $l=0$. This is demonstrated in fig. \ref{fig:ppp_vibro_modes_l}, where the modes with $l=1,2$ are plotted. One sees that unlike the $l=0$ case, there are no vibrational modes in the vicinity of the cusp point.

\subsection{Decay mode}
\label{ssec:decay_mode}

\begin{figure}[b!]
\centering
\begin{minipage}[c]{0.49\textwidth}
\includegraphics[scale=0.6]{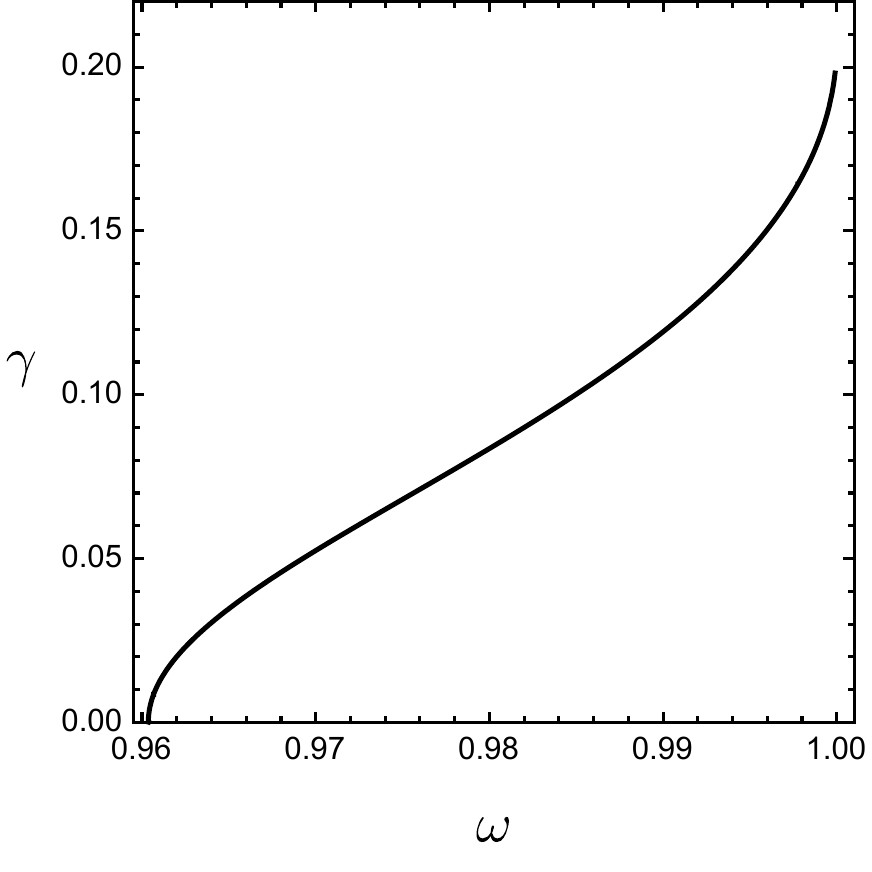}
\end{minipage}
\hfill
\begin{minipage}[c]{0.49\textwidth}
\includegraphics[scale=0.6]{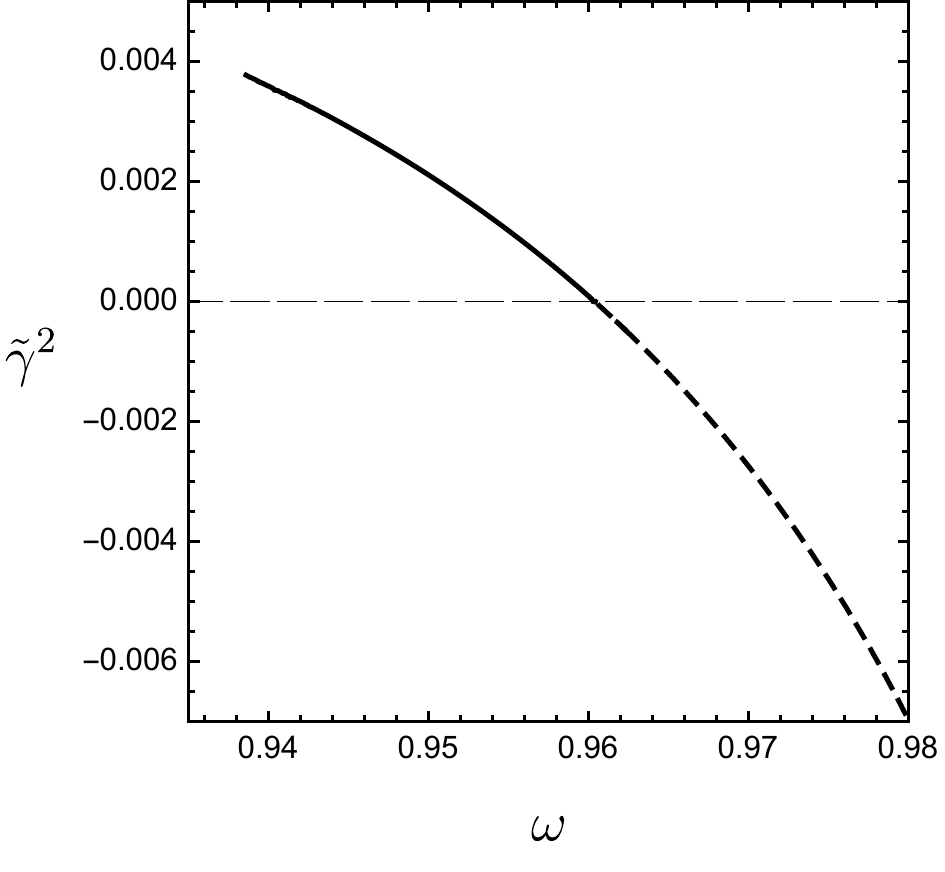}
\end{minipage}
\caption{ \textit{Left panel:} the decay rate of unstable Q-balls in the potential (\ref{appr:PPP}). \textit{Right panel:} the transition between the decay and vibrational modes, with $\tilde{\gamma}$ defined in eq. (\ref{ppp:gamma_tilda}); cf. fig. \ref{fig:ppp_vibro_modes}. }
\label{fig:decay_mode}
\end{figure}

The decay mode is captured by the following perturbation ansatz with $l=0$:
\begin{equation}\label{ppp:decay_mode_ansatz}
\psi(\vec{x},t)=\zeta(r)e^{\gamma t} \; ,
\end{equation}
where $\gamma$ is taken to be real and positive. Studying the condition under which this mode exists, one arrives at the Vakhitov-Kolokolov criterion $dQ/d\omega>0$ \cite{Vakhitov:1973} of the instability of Q-balls \cite{Alford:1987vs}. Substituting eq. (\ref{ppp:decay_mode_ansatz}) into eqs. (\ref{appr:psi}), we deduce
\begin{align}\label{ppp:decay_modes_equations}
\begin{split}
(\Delta_r-(\gamma+i\omega)^2+g(r))\zeta(r)+h(r)\zeta^\ast(r)=0 \; ,  \\
(\Delta_r-(\gamma-i\omega)^2+g(r))\zeta^\ast(r)+h(r)\zeta(r)=0 \; .
\end{split}
\end{align}
Imposing the boundary conditions
\begin{equation}
\zeta'(0)=0 \; , ~~~ \zeta(\infty)=0
\end{equation}
results in the solution
\begin{equation}\label{ppp:zeta_ansatz}
\zeta(r)  = \left\lbrace \begin{array}{l}  A \dfrac{\sin\left(\lambda r\right)}{r} \; , ~~~~  r<R \; , \\
\\
B\dfrac{e^{-\vartheta r}}{r} \; , ~~~~~~~ r> R \; , \end{array}\right.
\end{equation}
and the analogous expression for $\zeta^{\ast}$. Here we denote
\begin{align}\label{ppp:decay_frequncies}
\begin{split}
\lambda^2 & =\omega^2-\gamma^2-2 i \gamma \omega \; ,  \\
\vartheta^2 & = m^2+\gamma^2-\omega^2+2 i \gamma \omega \; ,
\end{split}
\end{align}
and $R$ is given in eq. (\ref{ppp:R}). The constants $A$, $B$, and $\lambda$ are found from a smooth matching of the two parts of the solution at $r=R$. For $\lambda$ we have
\begin{align}\label{ppp:det_eq_zero}
\begin{split}
& \left(|\lambda|^2-|\vartheta|^2+\vartheta_R\Lambda\right)\cos(2\lambda_R R) +\left(|\lambda|^2+|\vartheta|^2-\vartheta_R\Lambda\right)\cosh(2\lambda_I R)  \\
& + \left(2\vartheta_I\lambda_I+2\vartheta_R\lambda_R-\lambda_R\Lambda\right)\sin(2\lambda_R R)+\left(2\vartheta_R\lambda_I-2\vartheta_I\lambda_R-\lambda_I\Lambda\right)\sinh(2\lambda_I R)=0 \; ,
\end{split}
\end{align}
where the subscript $I$ ($R$) means imaginary (real) part, and $\Lambda$ is defined in eq. (\ref{ppp:L}).\footnote{In cases when the potential does not admit analytic solutions, eqs. (\ref{ppp:decay_modes_equations}) can be solved numerically; see, e.g., \cite{Levkov:2017paj}.} 

The solution of eq. (\ref{ppp:det_eq_zero}) is shown in fig. \ref{fig:decay_mode}. One observes that $\gamma \rightarrow 0$ as $\omega$ approaches the cusp point. It is easy to show that near the cusp
\begin{equation}
\gamma\sim\sqrt{\omega-\omega_c} \; .
\end{equation}
Comparing with eq. (\ref{ppp:near_cusp}), we see that the solution is continued analytically beyond the cusp point where it becomes the vibrational mode in the spectrum of Q-balls lying close to the stability bound. This is shown explicitly on the right panel of fig. \ref{fig:decay_mode} where we denote
\begin{equation}\label{ppp:gamma_tilda}
\tilde{\gamma}^2\equiv \gamma^2 ~~~ \text{for} ~~~ \omega<\omega_c \; , ~~~~~ \tilde{\gamma}^2\equiv -\gamma^2 ~~~ \text{for} ~~~ \omega\geq\omega_c \; .
\end{equation}

Note that, except for a vicinity of the point $\omega=1$, the decay rate $\gamma$ is small enough, which makes the perturbation theory with respect to $\gamma$ applicable almost everywhere along the unstable branch of Q-balls. For example, the form of the decay mode to the linear order in $\gamma$ is
\begin{equation}\label{ppp:psi_near_omega_cusp}
\psi e^{-\gamma t} \sim if+\gamma\frac{\partial f}{\partial\omega}+\mathcal{O}(\gamma^2) \; .
\end{equation}
In this expression, the first term represents the Goldstone mode corresponding to the global $U(1)$ symmetry of the theory.

\section{Thin-wall limit of Q-balls in Polynomial potential}
\label{sec:Poly}

\subsection{Ansatz}
\label{ssec:ansatz}

In general, a theory of the complex scalar field with a polynomial potential does not admit analytic Q-ball solutions in three spatial dimensions. Nevertheless, an analytical description of both the Q-balls and their excitations can be obtained in certain limits of parameters of the theory. One of these limits is governed by the so-called thin-wall approximation \cite{Coleman:1985ki}. In this regime, the properties of a Q-ball are well captured by a few quantities --- a distance $R$ to the wall separating interior and exterior regions of the configuration, and a magnitude $f_0$ of the scalar field in the interior. Hence, the thin-wall approximation allows us to reduce the full variational problem to a problem of finding a (conditional) extremum of a function depending on the finite amount of parameters.

In order to justify the transition from the description of a solution in terms of the fields to the description in terms of just a finite set of variables, a suitable field ansatz containing such variables must be adopted. The latter must be chosen in a way compatible with the equations of motion of the original theory. The thin-wall approximation implies the existence of a small parameter $\epsilon$ with respect to which one measures a degree of validity of the chosen ansatz. Perturbation theory built with $\epsilon$ will ensure the independence of the leading-order characteristics of the solution $R$ and $f_0$ of the details of the ansatz. 

In this section we consider the simplest bounded below polynomial potential of the sixth degree, which is conveniently parametrized as follows:
\begin{equation}\label{phi6:V}
V(|\phi|)=\left( \delta\left(|\phi|^2-v^2\right)^2+\omega_{\min}^2 \right)|\phi^2| \; ,
\end{equation}
where $\delta>0$. With this potential, the frequencies of nontopological solitons are confined in the region
\begin{equation}
\omega_{\min}<\omega<m=\sqrt{\omega_{\min}^2+\delta\,v^4} \; .
\end{equation}
The thin-wall approximation is applicable for stable Q-balls near the lower bound of this region, $\omega_{\min}>0$, and we expect 
\begin{equation}
\epsilon=\omega-\omega_{\min}
\end{equation}
to be an appropriate small parameter. The ansatz for the Q-balls in this limit can be chosen in a number of ways. For example, the form of the solution in ($1+1$) dimensions \cite{Belova:1994vd} suggests the following expression for the magnitude of the scalar field:
\begin{equation}
f(r)=\left\lbrace\begin{array}{l} f_0 \; , ~~~  r<R \; , \\
													f_0(\cosh^2 a(r-R)+b\sinh^2 a(r-R))^{-\frac{1}{2}} \; , ~~~ r\geq R  \\
						\end{array}\right.
\end{equation}
with $a,b>0$. It captures a more subtle structure of the solution, allowing us to compute, for example, a proper thickness of the wall, which is subleading to the radius of a Q-ball. However, for the purposes of studying perturbations on top of the Q-balls, it is enough to make the simplest choice possible,
\begin{equation}\label{phi6:Ansatz}
f(r)=f_0\theta\left(1-\frac{r}{R}\right) \; .
\end{equation}
Below we proceed with this form of the ansatz, by computing first $f_0$ and $R$ to the leading order in $\epsilon$, and then studying analytically the discrete spectrum of perturbations.

\subsection{Q-ball solution in the thin-wall regime}
\label{ssec:Sol_thin_wall}

With the ansatz (\ref{phi6:Ansatz}) applied, the global charge of a Q-ball is written as
\begin{equation}
Q= \frac{8}{3}\pi R^3 \omega f_{0}^{2} \; .
\end{equation}
Further, the energy of the soliton takes the form \cite{Coleman:1985ki},
\begin{equation}\label{phi6:E}
E=E_{\mathrm{surf}}+E_{\mathrm{vol}} \; ,
\end{equation}
where the surface energy of the wall is given by
\begin{equation}
E_{\mathrm{surf}}=\lim_{\omega \rightarrow \omega_{\mathrm{min}}}\int d^3 x \left((\nabla f)^2+V(f)-\omega^2 f^2\right) = 8 \pi R^2 \sqrt{\delta} v^4 \; ,
\end{equation}
and the energy of the Q-ball's interior is
\begin{equation}
E_{\mathrm{vol}}=\int d^3 x \left(\omega^2 + \omega_{\min}^{2}\right)f^2= \frac{4}{3}\pi R^3 \left(\omega^2+\omega_{\min}^2\right)f_{0}^{2} \; .
\end{equation}

The size $R$ of the Q-ball and its magnitude $f_0$ are found by minimizing the energy while keeping the charge fixed. To the leading order in $\epsilon$ this gives
\begin{equation}\label{phi6:f0,R}
f_0=v+\mathcal{O}(\epsilon) \; , ~~~~ R=\dfrac{\sqrt{\delta}v^2}{2\omega_{\min}}\dfrac{1}{\epsilon}+\mathcal{O}(1) \; .
\end{equation}
As expected, $R$ experiences a powerlike divergence as $\epsilon\rightarrow 0$. 

Eqs. (\ref{phi6:f0,R}) coincide with those obtained, e.g., in \cite{Mai:2012yc} after a suitable reparametrization of the potential (\ref{phi6:V}) is made. Note also that, as a cross-check, one can make sure that the well-known relation between the energy and the charge of a Q-ball,
\begin{equation}
\dfrac{dE}{d\omega}=\omega\dfrac{dQ}{d\omega}
\end{equation}
is satisfied upon substituting eqs. (\ref{phi6:f0,R}) into the expressions for $Q$ and $E$ and differentiating with respect to $\omega$.

\subsection{Vibrational modes in the thin-wall regime}
\label{ssec:Pert_thin_wall}

Proceeding as in the case of the piecewise parabolic potential studied before, we choose the ansatz for perturbations according to eq. (\ref{ppp:vibro_ansatz}) and substitute it into eqs. (\ref{appr:psi}) where the functions $h$, $g$ are taken as
\begin{align}
\begin{split}
h(r) & =-4\delta v^2 f^2(r)+6\delta f^4(r) \; ,  \\
g(r) & =m^2-8\delta v^2 f^2(r)+9\delta f^4(r) \; ,
\end{split}
\end{align}
and $f(r)$ is given in eqs. (\ref{phi6:Ansatz}) and (\ref{phi6:f0,R}). The resulting equations are solved analytically in the regions $r>R$ and $r<R$, and the corresponding solutions are subject to the boundary conditions (\ref{ppp:bc}) and the requirement of a smooth matching at $r=R$.

In the exterior of the Q-ball, $r>R$, a straightforward calculation gives
\begin{align}\label{phi6:r>R}
\begin{split}
\psi_1^{(l)}(r) & =C\frac{k_{l+\frac{1}{2}}\left(\sqrt{m^2-(\omega+\gamma)^2}r\right)}{r} \; ,  \\
\psi_2^{(l)}(r) & =D\frac{k_{l+\frac{1}{2}}\left(\sqrt{m^2-(\omega-\gamma)^2}r\right)}{r} \; ,
\end{split}
\end{align}
where 
\begin{equation}
k_{l+\frac{1}{2}}(x)=\sqrt{x}K_{l+\frac{1}{2}}(x) \; .
\end{equation}
Inside the Q-ball, $r<R$, the components $\psi$ and $\psi^*$ of the perturbation in eqs. (\ref{appr:psi}) cannot be disentangled. Let us denote $\Psi=(\psi_1,\psi_2)^T$. Then, in the matrix notation,
\begin{equation}\label{phi6:vib_eq_in_t_w}
\left(
\left(\Delta+\omega^2+\gamma^2-\alpha\right)\times \mathrm{1}_{2\times2}+2 \omega \gamma \sigma_3-2\beta\sigma_1
\right)\Psi=0 \; ,
\end{equation}
where $\sigma_{1,3}$ are the Pauli matrices and we introduced
\begin{equation}
\alpha= g(r<R) =m^2+\delta v^4 \; , ~~~~ \beta=\frac{h(r<R)}{2}=\delta v^4 \; .
\end{equation}
Let $U$ be the matrix diagonalizing the last two terms of the operator in the l.h.s. of eq. (\ref{phi6:vib_eq_in_t_w}). Then, introducing a vector $\Xi$ such that $\Psi=U\Xi$, one obtains separate equations for the components $\xi_1$ and $\xi_2$ of $\Xi$. Solving them yields
\begin{align}
\begin{split}
\xi_1^{(l)} & = A \frac{j_{l+\frac{1}{2}}\left( r \sqrt{\omega^2+\gamma^2-\alpha+2\sqrt{(\omega\gamma)^2+\beta^2}}\right)}{r} \; , \\
\xi_2^{(l)} & = B \frac{i_{l+\frac{1}{2}}\left( r \sqrt{\alpha+2\sqrt{(\omega\gamma)^2+\beta^2}-\omega^2-\gamma^2}\right)}{r} \; ,
\end{split}
\end{align}
where
\begin{align}
\begin{split}
j_{l+\frac{1}{2}}(x) & =\sqrt{x}J_{l+\frac{1}{2}}(x) \; , \\
i_{l+\frac{1}{2}}(x) & =\sqrt{x}I_{l+\frac{1}{2}}(x) \; ,
\end{split}
\end{align}
and $I_{l+\frac{1}{2}}(x)$ is the modified Bessel function of the first kind and of the order $l+1/2$. From this and an explicit form of the matrix $U$,
\begin{equation}
U=\left(\begin{array}{cc}
-\omega\gamma-\sqrt{\omega^2\gamma^2+\beta^2} & -\omega\gamma+\sqrt{\omega^2\gamma^2+\beta^2}\\
\beta & \beta 
\end{array}\right) \; ,
\end{equation}
one restores the components $\psi^{(l)}_1$ and $\psi^{(l)}_2$ of the perturbation.

The values of the constants $A$ to $D$ are found by matching smoothly the modes at the position of the wall $r=R$. The matching also determines an equation that allowable values of $\gamma$ must satisfy. The latter reads as follows:
\begin{align}\label{eq:det_t_w}
\begin{split}
& \left(-2\lambda_{+}\lambda_{-}\frac{k'_{l+\frac{1}{2}}(\lambda_{+}R)}{k_{l+\frac{1}{2}}(\lambda_{+}R)}\frac{k'_{l+\frac{1}{2}}(\lambda_{-}R)}{k_{l+\frac{1}{2}}(\lambda_{-}R)}\sqrt{\beta^2+\gamma^2\omega^2}+ \omega_{+}\lambda_{-}
\frac{j'_{l+\frac{1}{2}}(\omega_{+}R)}{j_{l+\frac{1}{2}}(\omega_{+}R)}\frac{k'_{l+\frac{1}{2}}(\lambda_{-}R)}{k_{l+\frac{1}{2}}(\lambda_{-}R)}\left(\gamma\omega+\sqrt{\beta^2+\gamma^2\omega^2}\right)\right.  \\
& ~~~~~~~~~~~~~~~~~~~~~~~~~~~~~~~~~~~~~~~~~~~~~~~~~~~~~~~~~~ +\left. \omega_{+}\lambda_{+}\frac{j'_{l+\frac{1}{2}}(\omega_{+}R)}{j_{l+\frac{1}{2}}(\omega_{+}R)}\frac{k'_{l+\frac{1}{2}}(\lambda_{+}R)}{k_{l+\frac{1}{2}}(\lambda_{+}R)}\left(-\gamma\omega+\sqrt{\beta^2+\gamma^2\omega^2}\right)  \right)  \\
& +\omega_{-}\frac{j'_{l+\frac{1}{2}}(\omega_{-}R)}{j_{l+\frac{1}{2}}(\omega_{-}R)}\left(-2\omega_{+}\frac{j'_{l+\frac{1}{2}}(\omega_{+}R)}{j_{l+\frac{1}{2}}(\omega_{+}R)}\sqrt{\beta^2+\gamma^2\omega^2}+\lambda_{-}\frac{k'_{l+\frac{1}{2}}(\lambda_{-}R)}{k_{l+\frac{1}{2}}(\lambda_{-}R)}\left(-\gamma\omega+\sqrt{\beta^2+\gamma^2\omega^2}\right)    \right.  \\
& ~~~~~~~~~~~~~~~~~~~~~~~~~~~~~~~~~~~~~~~~~~~~~~~~~~~~~~~~~~ +\left.\lambda_{+}\frac{k'_{l+\frac{1}{2}}(\lambda_{+}R)}{k_{l+\frac{1}{2}}(\lambda_{+}R)}\left(\gamma\omega+\sqrt{\beta^2+\gamma^2\omega^2}\right)\right)=0 \; ,
\end{split}
\end{align}
where
\begin{align}
\begin{split}
\omega_{\pm} & =\sqrt{2\sqrt{\omega^2\gamma^2+\beta^2}\pm\left(\omega^2+\gamma^2-\alpha\right)} \; , \\
\lambda_{\pm} & =\sqrt{m^2-(\omega\pm\gamma)^2} \; .
\end{split}
\end{align}

\begin{figure}[t]
\centering
\begin{minipage}[c]{0.49\textwidth}
\includegraphics[scale=0.3]{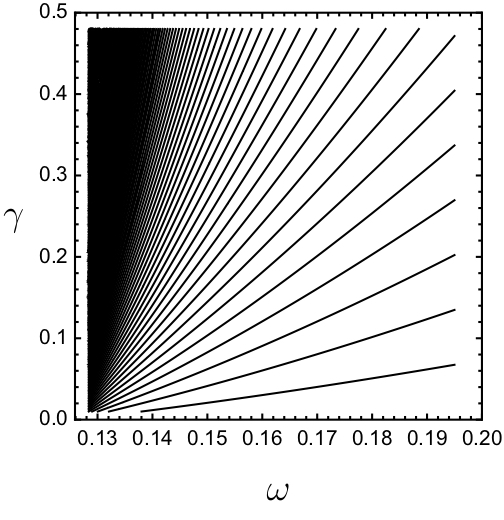}
\end{minipage}
\hfill
\begin{minipage}[c]{0.49\textwidth}
\includegraphics[scale=0.65]{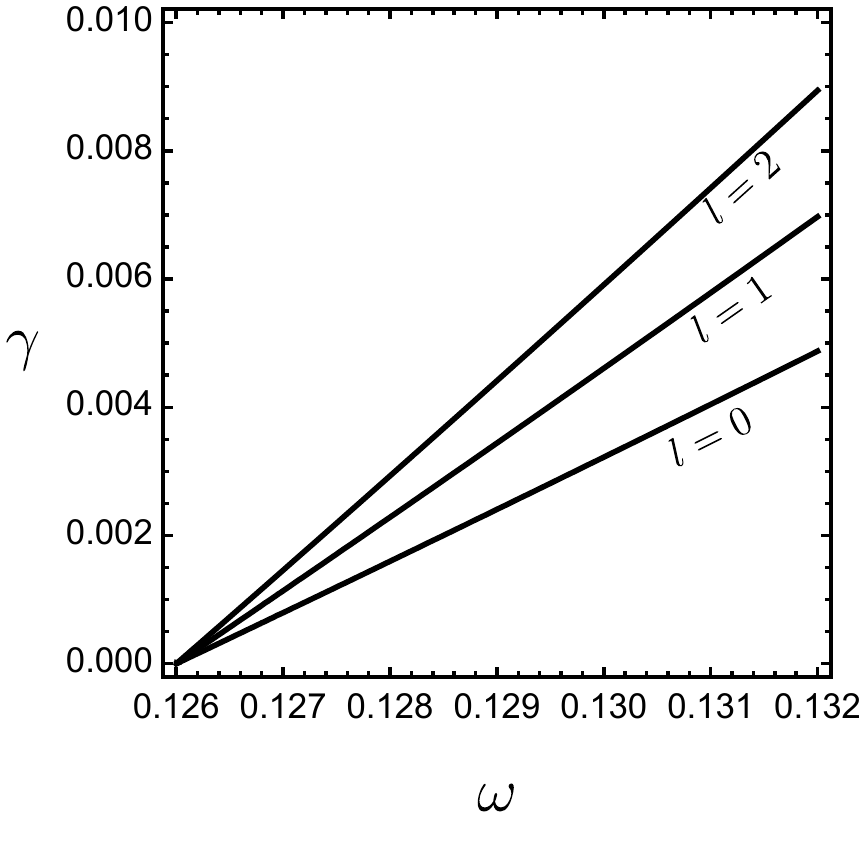}
\end{minipage}
\caption{ The spectrum of vibrational modes of stable Q-balls in the thin-wall approximation. The parameters of the potential (\ref{phi6:V}) are chosen as $\delta=1.5$, $v=0.9$, $\omega_{\min}=0.126$. All quantities are normalized to $m$. The left panel shows the full spectrum of the spherically symmetric modes, $l=0$. The right panel compares the modes with $n=1$ and with different orbital momenta. For illustrative purposes, only the region near $\omega_{\min}$ is shown. }
\label{fig:t_w_spectrum}
\end{figure}

Shown in fig. \ref{fig:t_w_spectrum} are the values of $\gamma$ solving eq. (\ref{eq:det_t_w}) at different Q-ball's frequencies and with a particular choice of the parameters of the potential. We see that the spectrum of vibrational modes around a Q-ball in the polynomial potential shows the same behavior as the one in the piecewise parabolic potential. In particular, all energy levels tend to zero as the lower bound $\omega=\omega_{\min}$ is approached, and near this bound the spectrum linearizes.

Eq. (\ref{eq:det_t_w}) allows significant simplification in the limit of large $R$. Using the properties of the special functions \cite{jeffrey2007table},
\begin{equation}
-\lim_{x\to\infty}\frac{k'_{l+\frac{1}{2}}(x)}{k_{l+\frac{1}{2}}(x)}=\lim_{x\to\infty}\frac{i'_{l+\frac{1}{2}}(x)}{i_{l+\frac{1}{2}}(x)}=1 \; , 
\end{equation}
we arrive at
\begin{equation}\label{phi6:det_t_w_lin}
\frac{j_{l+\frac{1}{2}}(\omega_{+}R)}{j'_{l+\frac{1}{2}}(\omega_{+}R)}=-\frac{\omega_{+}}{\omega_{-}}\frac{\gamma\omega\left(\lambda_{-}-\lambda_{+}\right)+\left(\lambda_{-}+\lambda_{+}+2\omega_{-}\right)\sqrt{\beta^2+\gamma^2\omega^2}}{\gamma\omega\left(-\lambda_{-}+\lambda_{+}\right)+\left(\left(\lambda_{-}+\lambda_{+}\right)+\frac{2\lambda_{-}\lambda_{+}}{\omega_{-}}\right)\sqrt{\beta^2+\gamma^2\omega^2}} \; .
\end{equation}
From fig. \ref{fig:t_w_spectrum} we see that at any given energy level
\begin{equation}
\gamma=k\epsilon \; .
\end{equation}
The coefficient $k$ can be found by substituting this into eq. (\ref{phi6:det_t_w_lin}). The result is
\begin{equation}
k_{n,l}=\frac{2 \omega_{\min}}{m}\mu_{n, l+\frac{1}{2}} \; , ~~~ n=0,1,2,...
\end{equation}
where $\mu_{n, l+\frac{1}{2}}$ denotes the $n$th zero of the Bessel function of the first kind and of the order $l+1/2$. 

Finally, let us evaluate the amount $N$ of discrete energy levels on top of a Q-ball in the thin-wall approximation. This amount is finite in view of the constraint
\begin{equation}
\gamma+\omega<m \; ,
\end{equation}
separating the bound states from the continuous spectrum. Using the asymptotics of the zeros of the Bessel function,
\begin{align}
\begin{split}
\mu_{n, l+\frac{1}{2}} & = \pi\left(n+\frac{l}{2}\right)+\mathcal{O}(n^{-1}) \; , ~~~ l ~~ \text{is fixed} \; , \\
\mu_{n, l+\frac{1}{2}} & = l + \mathcal{O}(l^{1/3}) \; , ~~~~~~~~~~~~~~~~ n ~~ \text{is fixed} \; ,
\end{split}
\end{align}
we have
\begin{equation}
N\sim \left(\dfrac{m-\omega_{\min}}{\epsilon}\dfrac{m}{\omega_{\min}}\right)^3\sim R^3 \; ,
\end{equation}
where, in the second estimation, we made use of eq. (\ref{phi6:f0,R}). We observe that the number of bound states is proportional to the internal volume of the Q-ball. One expects this result to hold in the general case. In particular, the similar counting shows that in the piecewise parabolic potential the number of bound states follows the same rule, $N\sim R^3$, where $R$ is identified with the matching radius (\ref{ppp:R}), provided that the size of the Q-ball is large enough.

\section{Conclusion and Acknowledgements}
\label{sec:Concl}

In this paper, our aim was to study properties of perturbations of (classically stable) Q-balls arising in theories of the complex scalar field in different setups. In choosing particular models for investigation, we were motivated by the possibility to perform the analytical treatment of both the background solitons and their excitations. Namely, we discussed the model with the piecewise parabolic potential containing the flat direction, which can approximate realistic theories relevant for cosmology. A complement to this case is the theories with potentials exhibiting a powerlike behavior at large fields. Here we considered the simplest bounded below potential of the sixth degree and focused on studying the thin-wall limit of Q-balls in this potential.

We found that the spectra of vibrations of Q-balls in our examples have some properties in common. This enables us to believe that their appearance is, in fact, insensitive to the particular scalar potential. We saw that the large Q-balls in the model with the flat direction support a bunch of soft modes with the frequencies $\omega\sim\gamma\rightarrow 0$, well below the mass of the boson in vacuum. The large Q-balls in the polynomial potential, on the other hand, do not contain soft modes due to the finiteness of the minimal frequency $\omega_{\min}$. Interestingly, the Q-balls with the nearly critical charges possess the vibrational mode which is related to the decay mode of the Q-clouds. It is important to note that the near-critical regime of these (in general, relativistic) solitons can be analyzed by the means of the perturbation theory with respect to the relative frequency $\gamma$ of an excitation. This result may be of some interest in studies of relativistic BS, for which the possibility of the analytical treatment is limited. 

The authors are grateful to D. Levkov and M. Smolyakov for useful discussions. The work was supported by the Russian Science Foundation Grant No. RSF 16-12-10-494.

\appendix
\section{More on mode expansion near the cusp point}
\label{sec:log}

Let us demonstrate that eqs. (\ref{ppp:psi_near_cusp_vibro}) represent the general form of linear perturbations of a Q-ball near the cusp point. To this end, we consider yet another potential admitting analytic Q-ball configurations. It reads as follows \cite{Marques:1976ri}:
\begin{equation}\label{log:potential}
V(|\phi|)=-m^2|\phi|^2\ln\left(\frac{\lambda |\phi|^2}{m^2}\right) \; ,
\end{equation}
where $m,\lambda>0$.\footnote{Note that because of the infinite mass of free quanta, the cusp point in this potential arises in the top-right corner of the $E(Q)$ plot. Hence, it sets an upper limit on the charge and energy of Q-balls, contrary to the common case when the cusp appears in the bottom-left and constrains the charge and energy from below. The former picture is more typical for BS (see, e.g., \cite{Friedberg:1986tp}).} The profile of a Q-ball is given by
\begin{equation}
f_0(r)=\frac{m}{\sqrt{\lambda}}e^{-\frac{(m r)^2}{2}-\frac{\omega^2}{2 m^2}+1} \; .
\end{equation}
The linear perturbations governed by the ansatz (\ref{ppp:vibro_ansatz}) can also be found analytically. With this ansatz applied, eqs. (\ref{ppp:vibro_pert_eqns}) become similar to the equation for the one-dimensional harmonic oscillator, and the perturbations $\psi_1$ and $\psi_2$ are eigenfunctions of that oscillator \cite{Marques:1976ri}. 

The oscillation rate near the cusp point is
\begin{equation}
\gamma = 2 \sqrt{\omega^2-\frac{m^2}{2}} \; ,
\end{equation}
and the corresponding vibrational mode reads as follows:
\begin{equation}
\psi_1=\left(1-\frac{\gamma\omega}{m^2}\right) f_0 \; , ~~~~ \psi_2=\left(1-\frac{\gamma\omega}{m^2}\right)\left(1+\frac{2\gamma\omega}{m^2}+\frac{\gamma^2}{m^2}\right) f_0
\end{equation}
up to an overall normalization constant. Extracting the linear order in $\gamma$, we obtain
\begin{equation}
\psi_1\sim f_0+\gamma \frac{\partial f_0}{\partial\omega}+\mathcal{O}\left(\gamma^2\right) \; , ~~~~ \psi_2\sim -f_0+\gamma \frac{\partial f_0}{\partial\omega}+\mathcal{O}\left(\gamma^2\right) \; ,
\end{equation}
in agreement with eq. (\ref{ppp:psi_near_cusp_vibro}). Thus, the perturbation theory with respect to the parameter $| \omega-\omega_c |^{1/2}$ works well for the potential (\ref{log:potential}).

\end{document}